\documentclass[fp,twocolumn]{jpsj3}
\usepackage{txfonts}
\usepackage{color}

\usepackage{ulem}

\title{Designing a Polymerized Phenalenyl Tessellation Molecule to Realize a Super-honeycomb Antiferromagnetic $S=3/2$ Spin System
}

\author{
Kenshin Komatsu$^1$, 
Naoki Morishita$^2$, 
Motoharu Kitatani$^1$, and 
Koichi Kusakabe$^1$
}

\inst{
$^1$Graduate School of Science, University of Hyogo, Kamigori, Hyogo 678-1297, Japan \\
$^2$Research and Development Directorate, Japan Aerospace Exploration Agency, Tsukuba, Ibaraki 305-8505, Japan
} 

\abst{
In a multiply hydrogenated polymer of phenalenyl tessellation molecules (PTMs), spatially overlapping zero modes appear, and three spin-aligned electron spins per PTM are generated through direct exchange interactions in the strongly correlated electron system. This interaction was used to design a two-dimensional (2D) $S = 3/2$ Heisenberg spin system on a honeycomb lattice. Simulations of the electronic structure using density functional theory with the Wannierization method revealed an array of nonbonding molecular orbitals (zero modes) in the hydrogenated nanographene structure. Our analysis of the onsite interaction strength indicated that each zero mode was half-filled with a spin-active electron owing to electron correlation effects. The low-energy subspace of the resulting zero mode-tight-binding model suggests the formation of a 2D antiferromagnetic $S = 3/2$ Heisenberg system with an entangled quantum spin ground state.
}

\begin{document}
\maketitle

\section{Introduction}
The introduction of measurement-based quantum computation\cite{PhysRevLett.86.5188, PhysRevLett.93.040503} has generated considerable interest in the realization of resource states in practical systems.\cite{PhysRevLett.97.110501, PhysRevLett.103.240504} This
has raised hopes that hardware systems could either possess or generate quantum entangled states, which are identified as topological cluster states in optical lattices\cite{PhysRevLett.105.013902, PhysRevLett.105.093601} and/or the spin-3/2 Affleck–Kennedy–Lieb–Tasaki (AKLT) state in spin systems.\cite{PhysRevLett.105.110502, PhysRevLett.106.070501} However, for the former system, implementation poses a challenge because measuring an individual atomic qubit is quite difficult. For the latter system, it is essential to achieve an accurately designed alignment of quantum spins.\cite{PhysRevLett.114.247204} Several attempts have been made to implement quantum spin-1 systems\cite{Morishita2021, PhysRevB.108.155423} and a spin-1/2 system on a star lattice\cite{doi:10.1021/acs.nanolett.3c04915} using carbon-based materials such as nanographene. A key design principle in these studies is the nonbonding molecular orbital (NBMO). The disjoint NBMO method involves creating NBMOs and connecting them in a disjoint manner, ensuring that the NBMOs remain as eigenstates of a hopping Hamiltonian.\cite{doi:10.1021/ja00456a010,Shima-Aoki,doi:10.1080/10587259308035713,Aoki-Imamura}

In a spin system, a qubit is typically characterized by a nuclear or electron spin. To facilitate measurable quantum oscillations, the strength and controllability of interspin interactions must be appropriately managed. While electron spins are often affected by their environments, leading to decoherence, we focus on an electron spin system in an isolated hydrocarbon molecule owing to the diverse strategies available for designing magnetic spin systems. This designed system should incorporate structural flexibility and maintain stability in its ideal isolated state. To achieve this, effective material design must use methods that generate strong magnetic interactions, which cannot be obtained solely through the disjoint NBMO method. Furthermore, the material should be developed with a strong correlation limit to ensure the presence of stable local spin moments at $S>1/2$.

We focus on hydrocarbon materials because each molecule can function as a 2D $S = 3/2$ spin system. Using polymerized phenalenyl tessellation for molecular design enables us to create desirable connections among $\pi$ electron orbitals.\cite{MORISHITA_PLA_2021} A key strategy involves designing zero modes, which are NBMOs with topological origins. These zero modes adhere to the zero-sum rule, where the wave function amplitudes on adjacent sites sum to zero. 
The emergence of $S = 3/2$ spins is attributed to the presence of three zero modes that spatially overlap and are spin-aligned through ferromagnetic direct exchange interactions.

Zero modes can arise from both edges and point defects. For instance, triangulene is a molecule where edge-related zero modes appear multiple times. Indeed, triangulene, an alternating hydrocarbon, has been discussed in relation to its potential magnetism in polymers. \cite{doi:10.1021/acs.nanolett.9b01773,PhysRevB.108.155423,doi:10.1021/acs.nanolett.3c04915} In this context, several zero modes appear at the edges of the $\pi$ network, indicating the existence of a type of degenerate edge zero mode. Conversely, this paper will focus on zero modes resulting from point defects in the $\pi$ network. Our group has explored cases where point-defect zero modes appear in the eigenstates of a defined tight-binding model.

To ensure the viability of the material solutions, we used a renormalization approach. Density functional theory (DFT) simulations \cite{Hohenberg-Kohn, Kohn-Sham} and the Wannierization method \cite{Marzari-Vanderbilt} were used to assess the structures of the hydrocarbon materials and to verify their stability. A low-energy effective model was developed using the Wannierization method, and the evaluation of effective inter-electron interactions suggested the formation of an antiferromagnetic $S = 3/2$ spin system.

The remainder of this paper is organized as follows. The next section introduces the design principles for nanographene developed through phenalenyl tessellation molecules (PTMs) in our previous studies. It also discusses the physical properties of nanographene of interest and the computational methods used in this research. Section 3 presents the numerical results of the proposed system along with a thorough analysis. Finally, Section 4 provides the conclusions.

\section{Molecular Design and Confirmation Methods}

\subsection{$\pi$-electron Model of Nanographene}

Graphene is a two-dimensional honeycomb structure made up of carbon atoms. It is important to note that this honeycomb lattice consists of two sublattices. Nanographene refers to a group of materials that contain a fragment of graphene along with derivatives of that fragment.\cite{Enoki_Ando}  

A notable example of nanographene can be found in molecules where carbon atoms are arranged in a polycyclic aromatic structure. In particular, we will focus on molecules that consist of conjugated benzene rings.\cite{Krygowski} This arrangement can be represented as a graph, with each vertex corresponding to a carbon atom and the edges indicating the covalent $sp^2$ bonds.

At the edges of the graph, carbon atoms with unsatisfied valence may appear. To stabilize these reactive carbon sites, which could potentially have dangling bonds, it is common to terminate these unsatisfied bonds with hydrogen atoms. In this scenario, a nanographene molecule is classified as an “alternant hydrocarbon.” A stable alternant hydrocarbon molecule is characterized by two sets of carbon atoms: one set consists of starred carbon atoms, while the other contains unstarred carbon atoms. In physics, these correspond to A sub-lattice sites and B sub-lattice sites.

In each carbon atom of an alternant hydrocarbon, there exists one $\pi$ orbital, which we consider as a site for capturing an electron. The $\pi$ electrons that occupy these $\pi$ orbitals contribute to $\pi$ bonding. The system of $\pi$-bonded electrons is governed by the hopping motion of electrons, allowing the formation of delocalized molecular orbitals.

An effective way to describe this delocalization is through an electron Hamiltonian, presented in matrix form, where its eigenvectors correspond to energy eigenstate wavefunctions. Specifically, electron hopping, also known as electron transfer, is typically well-represented by the tight-binding model (TBM) of electrons. In alternant hydrocarbon molecules, electrons form strong $\sigma$-bonding bands as well as $\pi$-bands. When we analyze the $\pi$ electron system separately from the all-electron system, the model consists solely of $\pi$ orbitals. This approach is commonly referred to as the single- orbital tight-binding model (so-TBM).

If we assume that the electron transfer Hamiltonian contributes only to nearest-neighbor pairs of $\pi$ sites and that the transfer amplitudes are uniform, then we can express the so- TBM in the following mathematical form of second quantization.

\begin{equation}
    H_{\rm so-TBM} = -t_\pi \sum_{\langle i,j \rangle, \sigma}  
    \left(c^\dagger_{i,\sigma} c_{j,\sigma}+{\rm H.c.}\right), 
\end{equation}

Each $\pi$-site is represented by an integer $i$, while $\sigma$ denotes the electron spin. The
summation over $\langle i,j \rangle$ is taken for nearest-neighbor site pairs. In this context, $c_{i,\sigma}$ and $c^\dagger_{i,\sigma}$ signify the annihilation and creation operators, respectively, for an electron with $\sigma \in \{ \uparrow,\downarrow \}$ at site $i$. Nearest neighbor sites are linked by a transfer term by $t_\pi$ , which is assumed to be a constant for any pair of nearest neighbor sites $i$ and $j$. This so-TBM has served as a foundational framework for theoretical studies of carbon nanotubes and
graphene.\cite{Saito_Dresselhaus,Enoki_Ando} 

When electron transfer is depicted as edges or bonds in the graph of an alternant hydrocarbon molecule, and non-zero transfer occurs only between each pair of nearest- neighbor sites, the resulting graph forms a bipartite structure. In this structure, every edge connects a starred carbon site (an A site) to an unstarred carbon site (a B site). Consequently, the graph of nanographene becomes bipartite.

The so-TBM was used to investigate the physical properties of nanographene, particularly concerning its edges and defects. It is known that peculiar non-bonding orbitals and localized electronic states can occur at the edges of nanographene. These characteristic non-bonding orbitals have a distinct nature, with energies close to zero positioned between bonding $\pi$ states with negative energy and anti-bonding $\pi$ states with positive energy.

The theoretical investigation of non-bonding edge states has been systematically performed by treating these states as zero modes of so-TBM. Various methods have been developed to theoretically determine the emergence of edge states using so-TBM.\cite{Enoki_Ando} 

The zero mode of so-TBM corresponds to a zero-energy eigenstate of $H_{\rm so-TBM}$. 
\begin{equation} 
    H_{\rm so-TBM}\; | \psi_{ZM, I} \rangle = 0 \times |\psi_{ZM, I} \rangle.
\end{equation} 
Here, $| \psi_{ZM, I} \rangle$ defines the kernel of $H_{\rm so-TBM}$. When introducing a nanographene structure, multiple instances of $| \psi_{ZM, I} \rangle$ ($I=1,\cdots,N_{ZM}$) can arise, where $N_{ZM}$ represents the number of zero modes.

Around topological point defects in nanographene, similar electronic states exhibiting non-bonding characteristics may also appear. However, only a few established rules regarding the spectrum of $H_{\rm so-TBM}$ are known, specifically concerning the generation of defect-originated zero modes. This issue arises because detailed boundary conditions at the edges of nanographene, away from the defect, must be specified to confirm the creation of zero modes. The rules for the PTMs discovered by Morishita et al. provide a clear solution to this problem.\cite{Morishita2016, Morishita2019, MORISHITA_PLA_2021, Morishita2021} This set of rules also captures the characteristics of so-TBM.

We will present the PTM rules in the next section. These rules offer a considerable advantage for designing magnetic molecules. By following the PTM guidelines, we can easily identify graphs with multiple zero modes. Consequently, we can conclude that zero-mode magnetism for $S = 3/2$ Heisenberg models arises from the arrangement of multiple zero modes.

\subsection{PTM Rules for the $\pi$-network}

The phenalenyl molecule is a polycyclic alternant hydrocarbon composed of three benzene rings. To create hydrocarbon systems with properly arranged zero modes, we chose to generate a polymerized phenalenyltessellation molecule (poly-PTM). \cite{Morishita2016, Morishita2019, MORISHITA_PLA_2021, Morishita2021} We can draw diagrams that depict the positions of the $\pi$ sites, each contributing a $\pi$ electron, along with their bond connections.

\begin{figure}[htbp]
    \centering
    \includegraphics[width=8cm]{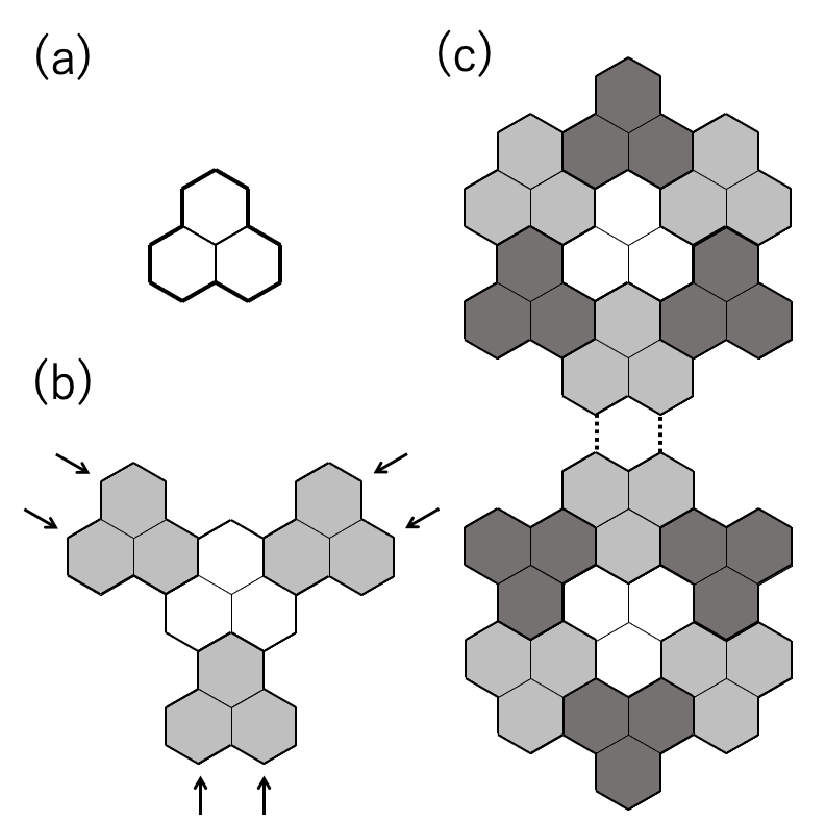}
    \caption{
    (a)A $\beta$-Phenalenyl unit (A $\beta$-PU). (b) Example of an ($\beta$-)phenalenyltessellation molecule (PTM), with the double zigzag corners (DZCs) indicated by arrows. (c) Example of a polymerized PTM (poly-PTM), where the $\alpha$-PTM (lower) and $\beta$-PTM (upper) are connected at the DZCs as indicated by dashed lines.
    }
    \label{FigPTM}
\end{figure}

As depicted in Fig.~\ref{FigPTM}(a), the carbon network is diagrammatically created only from phenalenyl units (PUs). Any polyPTM is formed by linking multiple PTMs through bond connections. However, a single PTM is not created by chemically bonding phenalenyl molecules together. Instead, a PTM represents a unique carbon network, and its skeleton diagram features PUs distinguished by different colors in Fig.~\ref{FigPTM}(b). In one PTM, two neighboring PUs may share a bond. For instance, a PTM can include a phenalenyl molecule but will not incorporate the triangulene.

After several PTMs are prepared, they can be connected to construct a poly-PTM (Fig. 1(c)), with the $\alpha$-PTM and $\beta$-PTM defined based on the “direction” of the phenalenyl structure. \cite{MORISHITA_PLA_2021} The three benzene rings in a PU form a triangle; an $\alpha$-PTM features only
PU triangles directed downwards, while a $\beta$-PTM consists solely of PU triangles directed upwards.(Fig.~\ref{FigPTM}(a)) The only inter-PTM connections occur between an $\alpha$-PTM and a $\beta$-PTM, treated as chemical bonds. To stabilize the structure of the poly-PTM, each edge of the entire carbon network needs to be terminated with hydrogen.

To create a 2D honeycomb network of PTMs, we can use triangular PTMs as the fundamental building blocks. The lattice features a rhombic unit cell with two “super” sites: $\alpha$ and $\beta$. The $\alpha$-PTM is defined according to whether the triangle (note that this corresponds to the directions triangle of PUs) is directed upward or downward. In contrast, the $\beta$-PTM is the mirror image of the $\alpha$-PTM. The $\alpha$-PTMs and $\beta$-PTMs are arranged at two sub-lattice sites in a super-honeycomb lattice. Neighboring carbon sites are
subsequently connected to form a 2D honeycomb network of PTMs, referred to as poly- PTM.

Creating additional vacancies in the poly-PTM can yield unique electronic
characteristics. When an atomic vacancy is introduced at the center of a PU, a localized
electronic state $| \psi_{ZM, I} \rangle$ with a nonbonding nature appears. Notably, this localized state is
centered at the vacancy, and its eigenenergy is found to be zero.
Therefore, the vacancy-centered localized mode can be classified as a zero mode. 
It may
appear that the mismatch between the numbers of sublattice sites, $N_A$ and $N_B$, in the bipartite lattice is responsible for the creation of this localized zero mode. However, this localized zero mode can also arise when $N_A = N_B$.
Because the number of vacancies equals the number of localized zero modes, we can achieve multiple zero modes.

The zero mode $| \psi_{ZM, I} \rangle$ localized around the center of the PU can be generated by adding
an electron site above the central site. These additional electronic sites should be
considered as sublattice sites. Then, $N_A$ and $N_B$ will be modified, and this results in the appearance of zero modes. The zero modes resulting from atomic defects and those induced by the addition of electronic sites are similar. For a more detailed discussion of the relationship between the two, please refer to the Supplementary material.\cite{SM} 
We will use this method of zero-mode creation to design quantum anti-ferromagnetic nanographene.

\subsection{$\pi$-network Design of $S=3/2$ Spin Systems}

We have demonstrated the construction of a one-dimensional $S=1$ Heisenberg model, specifically an antiferromagnetic $S=1$ chain, using a design method based on PTMs with localized zero modes.\cite{Morishita2021}  In this example, each PTM building block in the poly-PTM chain contains two vacancies, which induce two localized zero modes around the vacancies in each PTM, effectively forming an $S=1$ Heisenberg model. In each PTM, the two electron spins align ferromagnetically, resulting in a net $S=1$ spin for each PTM. Between adjacent PTMs, an antiferromagnetic effective interaction emerges between these $S=1$ spins.

Guided by this design method, we propose a nanographene structure where each PTM contains three vacancies, which is expected to realize an $S=3/2$ honeycomb Heisenberg model. To introduce three defects, each PTM must be composed of three or more PUs. Furthermore, to construct a honeycomb lattice from PTMs, the overall poly-PTM is designed to have two different PTMs, denoted as $\alpha$-PTM and $\beta$-PTM, in each rhombic unit cell. This requirement is essential for this study. The $\beta$-PTM with three vacancies is obtained by mirroring the $\alpha$-PTM with three vacancies. 

\begin{figure}[htbp]
    \centering
    \includegraphics[width=8cm]{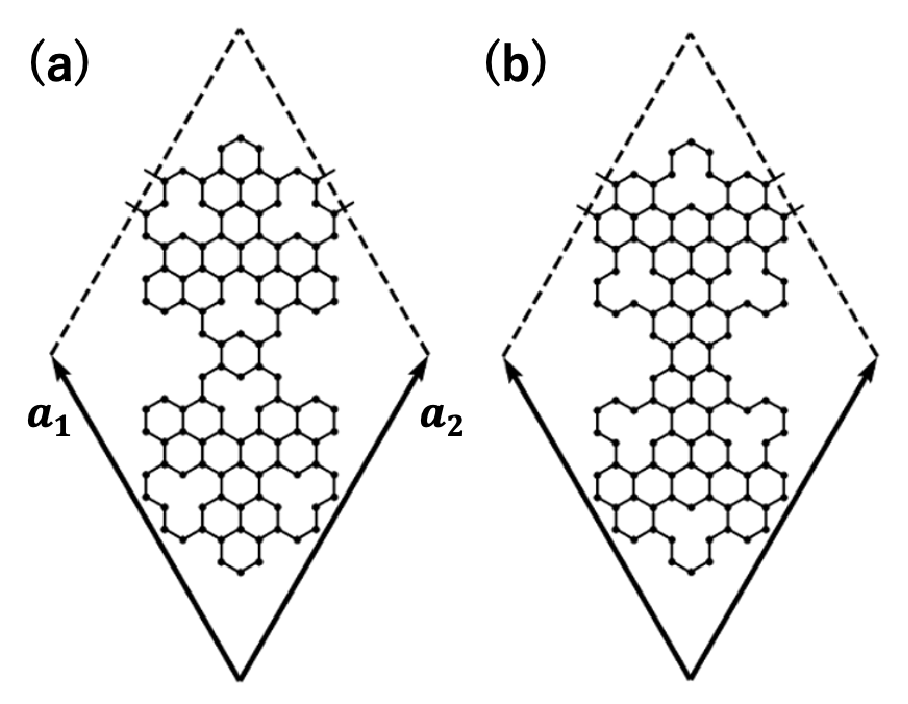}
    \caption{Designed 2D carbon frameworks:  
    (a)VANG-A nanographene and (b)VANG-B nanographene. ${\bf a}_1$ and ${\bf a}_2$ are the lattice vectors.}
    \label{f1}
\end{figure}

Fig.~\ref{f1} shows two examples of poly-PTMs arranged in a honeycomb lattice of zero modes. 
In Figs.~\ref{f1}~(a) and (b), $\alpha$-PTM represented by a molecular structure comprising 7 PUs. Three vacancies are introduced at the centers of three PUs. Accordingly, this $\alpha$-PTM with vacancies is referred to as a vacancy-created armchair nanographene molecule A (VANG-A) for the structure in Fig.~\ref{f1}~(a) and VANG-B for the structure in Fig.~\ref{f1}~(b). These poly-PTMs are derived from VANG-A (or VANG-B) in VANG-A nanographene (or VANG-B nanographene) shown in Fig.~\ref{f1}~(a) (or (b)). 
In both structures, three localized zero modes are induced in each PTM. These three localized zero modes give rise to a spin $S=3/2$ in each PTM. Furthermore, antiferromagnetic interactions are expected to emerge between the aligned spins in each PTM, leading to the realization of an $S=3/2$ honeycomb Heisenberg model. The proposed method for designing nanographenes using PTMs enables the proposition of diverse nanographene structures beyond those presented in this study. However, this research focuses on these two specific structures to compare the differences in magnetic properties that arise from distinct vacancy arrangements in the same fundamental skeleton. Therefore, our discussion will primarily focus on these two structures.

\begin{figure*}[t]
    \centering
    \includegraphics[width=18cm]{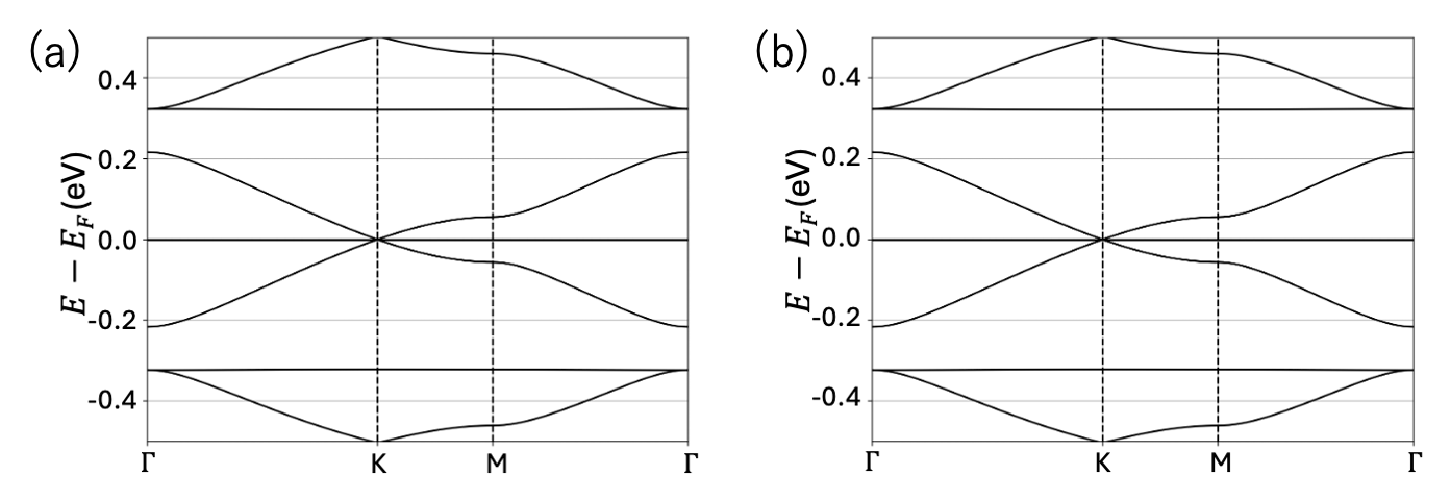}
    \caption{Band structures of the so-TBM for (a) VANG-A nanographene and (b) VANG-B nanographene. Both structures exhibit a sixfold degeneracy of zero modes. The two band structures are nearly identical up to energy levels around $\pm 0.7$ eV .}
    \label{f3}
\end{figure*}

Subsequently, We show the band structure of the so-TBM for these structures in Fig.~\ref{f3}. The localized zero modes at $E_F=0$ can be confirmed from the band structures obtained for each molecular framework in Fig.~\ref{f1} because flat bands with no dispersion appear at zero energy. Both molecular frameworks exhibit six flat bands at zero energy, which correspond to the localized zero mode for each of the six atomic vacancies in a unit cell. This characteristic can be predicted by the super-zero-sum rule.\cite{MORISHITA_PLA_2021} A Dirac point appears at the K (or K') point where a conduction band and valence band touch.

\subsection{Analysis of the $\pi$-network Using the Hubbard Model}

At each $\pi$ site, inter-electron interaction exists, namely, the on-site Hubbard interaction $U_{\pi}$, for electrons occupying the same $\pi$ orbital. Owing to $U_{\pi}>0$, we may consider a single-orbital Hubbard model (so-HM):
\begin{equation}
    H_{\rm so-HM} = -t_\pi \sum_{\langle i,j \rangle, \sigma}  \left(c^\dagger_{i,\sigma} c_{j,\sigma}+{\rm H.c.}\right)  + U_\pi \sum_i n_{i,\uparrow} n_{i,\downarrow}.
    \label{so-HM}
\end{equation}
By adopting the so-HM and referencing two relevant theorems of the Hubbard model, \cite{Lieb-Loss-McCann,Lieb} we can obtain a fundamental understanding of the correlation effects and magnetism in PTM structures. 

In the so-HM, where orbital degeneracy appears in the so-TBM with U=0, unconventional magnetism can appear. There, a type of ferromagnetic effective exchange interaction may arise from the Hubbard interaction. \cite{doi:10.1143/JPSJ.61.1165,PhysRevLett.72.144} 
Notably, the Lieb theorem, applicable to bipartite lattices, states that the total spin of the ground state of so-HM with $U>0$ is given by $S_{tot} = |N_A-N_B|/2$.\cite{Lieb}
Additionally, a discussion using perturbation theory was performed for a system in which orbital degeneracy occurs owing to edge/zero modes.\cite{doi:10.1021/acs.nanolett.9b01773,PhysRevB.108.155423} In this section, we discuss magnetism in poly-PTM adopting so-HM and also highlight its limitations.

As previously discussed, all $\pi$ bands can be classified into bonding, antibonding, and nonbonding bands. When the number of electrons equals the number of sites in the so-TBM, the system becomes charge-neutral. Under stable conditions, the lower bonding bands are completely filled by electrons. The zero mode, which is a nonbonding mode at $E_F$, should appear exactly between the bonding and antibonding bands and become half-filled. This is consistent with the chemical picture of NBMOs in alternant hydrocarbon systems.

Based on the uniform density theorem,\cite{Lieb-Loss-McCann} we can rule out the possibility of charge segregation and/or charge density waves in the lattice of zero modes. Each zero mode is localized around one of the vacancy sites. If any of the localized zero modes is not half-filled, the total occupation of the $\pi$ orbital would be non-uniform, contradicting the uniform density theorem. Thus, the appearance of half-filled zero modes is implied when we consider $U_{\pi}>0$ as a weak but finite perturbation of the so-TBM. The zero modes of the so-HM can also be expected to be half-filled. When the interaction $U_{\pi}$ is larger than $t_\pi$, because this perturbative picture is no longer valid, determining the occupation of zero modes becomes not easy, although the occupation of $\pi$ sites is uniform. 

A half-filled zero mode can exhibit a strong electron correlation effect, potentially leading to interesting magnetic behavior. Here, we consider the magnetism of the Hubbard model applied to the $\pi$-network of a poly-PTM. Let us consider the $\alpha$-PTM, which is one of the poly-PTMs shown in Fig.~\ref{f1}. Each skeleton diagram of the PTM is bipartite, meaning that we can divide the $\pi$ sites into A-sites and B-sites. The magnetic ground state can then be understood by applying the Lieb theorem to the repulsive Hubbard model with $U_\pi>0$.\cite{Lieb} 
%%This sentence is excluded! %%This approach is not necessarily strong enough to draw conclusions on the quality of the material design, but it can help with clarifying the physical mechanism of magnetism in nanographene.
To determine the total spin $S_{tot}$   of the ground state of the so-HM, we need to count the sublattice site numbers, as summarized in Table~I.

\begin{table}
    \centering
    \begin{tabular}{lcccccccc} \hline
   Structure & $N_A$ & $N_B$ & $|N_A-N_B|$ & $N_{\rm LZM}$ & $N_{\rm DZM}$ & $N_{\rm ZM}$ &  $S_{\rm tot}$ \\ \hline
    VANG    & 30 & 28 & 2 & 3 & 1 & 4 & 1 \\
    Dimer    & 58 & 58 & 0 & 6 & 0 & 6 & 0 \\
    poly-PTM & $58N_c$ & $58N_c$ & 0 & $6N_c$ & 2 & $6N_c+2$ & 0 \\
   \hline
   \end{tabular}
\label{t0}
\caption{
Number of sites, zero modes, and total spin of the ground state of the so-HM for each PTM structure. The VANG refers to the vacancy-created armchair nanographene molecule with three vacancy sites, and the dimer is composed of two VANGs. Here, Poly-PTM corresponds to the periodic system of VANGs with $N_c$ unit cells (i.e., $N_c$ corresponds to the number of k-points in the band calculation). $N_A$ and $N_B$ denote the number of sublattice sites. $N_{\rm LZM}$, $N_{\rm DZM}$, $N_{\rm ZM}$ represent the numbers of vacancy-centered localized zero modes, delocalized zero modes, and total zero modes, respectively. In the so-HM, the Lieb theorem determines the total spin $S_{\rm tot}$ of the ground state. }
\end{table}

If we define B-sites as including the center site of a PU, the number of B-sites $N_B$ of VANG is two less than the number of A-sites $N_A$ because the B-sites have three vacancies (Table~I).  Consequently, because the ground state total spin $S_{\rm tot}$ is proportional to the difference $|N_A-N_B|$, it becomes $S=|N_A-N_B|/2 = 1$ for one $\alpha$-PTM given by VANG. Because one $\alpha$-PTM has three localized zero modes and one delocalized zero mode,\cite{Morishita2019} a natural explanation for the $S=1$ ground state is that three of the four zero modes should have parallel electron spins, while the fourth electron has an antiparallel spin. Notably, the localized zero modes have nonzero amplitudes only on A sites, while the delocalized zero mode is on B sites.

If we consider a dimer comprising an $\alpha$-PTM and a $\beta$-PTM and construct a so-HM for the dimer, we can observe that six localized zero modes emerge, and there are no delocalized zero modes in the dimer. This is because the localized zero modes of PTMs are unaffected by the inter-PTM transfer in the so-TBM, which does not contain next-nearest-neighbor transfer integrals. In contrast, the number of delocalized zero modes is influenced and determined by the inter-PTM transfer. Two delocalized zero modes of two PTMs hybridize and form weak bonding and anti-bonding states that extend away from zero energy. 

According to the Lieb theorem,the total spin then becomes zero for the dimer. This result can be explained consistently in terms of perturbation theory. Three zero modes in one PTM can be represented by three orthogonal wave functions. Each localized orbital is centered at one of the vacancies. Any pair of the three localized orbitals has overlaps at several carbon $\pi$ sites. This results in a finite positive matrix element of the Hubbard interaction for any two electrons with antiparallel spins occupying these orbitals. The spin states of the three electrons can be either $S=1/2$ or $S=3/2$. Any $S=1/2$ state is affected by the Hubbard interaction, resulting in an energy increase by $U_{\pi}$.

In contrast, the $S=3/2$ state is formed only from parallel spins. There is no energy increase by the on-site Hubbard interaction owing to the Pauli principle. Therefore, the first-order perturbation of three-electron states in the three localized zero modes implies that the lowest energy state has a parallel spin $S=3/2$. As a result, electrons filling these three localized states of one PTM tend to have parallel spins, forming an $S=3/2$ spin state. The inter-PTM magnetic interaction becomes antiferromagnetic because of the possible kinetic exchange between two PTMs , resulting in a full ground state of the dimer being a singlet state.

%%This sentence is excluded! %%As an example, consider a dimer comprising one $\alpha$-PTM and one $\beta$-PTM. A small inter-PTM transfer makes the two delocalized zero modes in each PTM hybridize to form a bonding state and antibonding state. These hybridized states are lifted along the energy axis from the zero energy. Then, we can adopt the perturbation approach with $U$ as the perturbation Hamiltonian and only consider electrons in the localized zero modes. Then, we can explain the spin-polarization mechanism hidden in the $S=0$ ground state of the PTM dimer by using the Lieb theorem.

%%This sentence is excluded! %%Suppose that the inter-PTM bonds are another perturbation in a poly-PTM skeleton. Among the three localized zero modes on one PTM, the first-order perturbation in the Hubbard interaction results in three parallel spins as discussed above. Thus, two $S=3/2$ spins in these nanographene fragments are coupled antiferromagnetically to create a total spin state of $S=0$. Therefore, a small inter-PTM transfer creates an antiferromagnetic interaction. However, it cannot be the first order because the inclusion of the inter-PTM transfer keeps each zero mode as an eigenstate of the so-TBM. Therefore, the antiferromagnetic effective interaction is a second- or higher-order transfer term.

Thus, we may conclude that the spins of localized zero modes in one PTM tend to align ferromagnetically. Meanwhile, inter-PTM interactions should be antiferromagnetic to maintain consistency with the Lieb
theorem. This argument is fundamentally valid for poly-PTM as well. Thus, these zero modes are expected to induce magnetic properties in nanographene.

However, the abovementioned evaluation using the so-HM
(Eq.~(\ref{so-HM})) is insufficient for accurately assessing real hydrocarbon systems. One key reason is the influence of the long-range component of the Coulomb interaction. Even when appearing as a screened Coulomb interaction, off-diagonal components, such as intersite interactions and contributions from exchange terms, are essential for evaluating real materials. Because these components are not explicitly accounted for in the so-HM, it may not adequately capture the effective magnetic interactions of real materials.

 Another reason is that, in the so-HM, magnetic interactions among electrons accommodated by a zero mode may only occur via induced spin polarization in the other bonding $\pi$ bands. Therefore, they can appear only as higher-order perturbation terms, which have a weak influence.  
Meanwhile, in a real molecule, long-range hoppings will generate inter-zero mode hopping terms and inter-zero mode kinetic exchange terms, which induce hybridization and magnetic interactions among designed zero modes in poly-PTM. 
Thus, an alternative modeling approach is required, starting from the designed hydrocarbon system, to obtain stronger conclusions on the resultant magnetism than can be obtained from the so-HM of the $\pi$ electron model.

\section{Material Solutions by DFT-based Modeling}
\subsection{Molecular Model and Stable Structure Calculation Using DFT}
\begin{figure}[htbp]
    \centering
    \includegraphics[width=8cm]{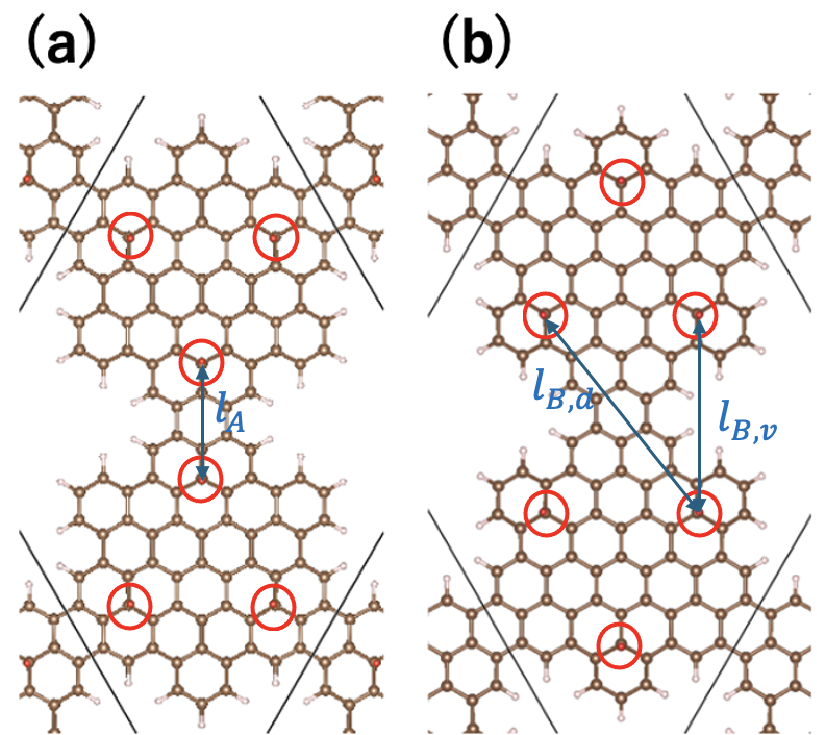}
    \caption{(Color online) Molecular models of (a)VANG-A nanographene, and (b)VANG-B nanographene. The red circles indicate the surface hydrogens. In VANG-A, the hydrogen distance is denoted as $l_A$, while in VANG-B, the vertical hydrogen distance is $l_{B,v}$ and the diagonal hydrogen distance is $l_{B,d}$ .}
    \label{f2}
\end{figure}

DFT simulations of the electronic structure allow us to progress from evaluating the stability of the material structure to determining the effective electron model. With this approach, we convert the skeleton diagram into the appropriate atomic configuration of the hydrocarbons. The stability of the resulting material can then be tested by simulations to optimize the structure. A material that is stable according to DFT simulations can confirm the expected appearance of zero modes. Deriving an effective electron model can elucidate whether the material realizes the target spin system. Therefore, we created a molecular model of hydrogenated nanographene based on the molecular skeleton (Fig. \ref{f2}) and calculated the electronic state using DFT simulations to determine the dispersion relation, Wannier functions, and electron transfer Hamiltonian.

In the molecular model of hydrogenated nanographene, several key choices and methods are necessary to induce the localized zero modes that appear in the carbon framework. A localized zero mode centered at the center site of a PU can be realized in two ways. One is to create a hydrogenated vacancy. The triply hydrogenated vacancy ({\it i.e.}, $V_{111}$ ) is a suitable solution\cite{Ziatdinov} but its creation severely deforms the graphene framework. The other approach is to create a vacancy in the graphitic carbon system by facilitating on-top hydrogen adsorption.\cite{Morishita2021} We opted for on-top hydrogen adsorption because it causes less distortion to the graphene structure and should be more manageable in practical experiments. With the abovementioned considerations in mind, the key steps of our design method for hydrogenated nanographene are as follows:

\begin{enumerate}
\item Carbon atoms are positioned at the vertices of the skeleton diagram of the poly-PTM shown in Fig.~\ref{f1}.
\item Carbon atoms are positioned in the vacancies of the skeleton under the assumption that on-top hydrogen adsorption occurs at these carbon sites.
\item The edges of the nanographene network are assumed to terminate in hydrogen atoms.
\end{enumerate}

Edge termination of the designed hydrogenated nanographene is assumed to be done by hydrogen atoms. This condition could potentially cause a steric hindrance effect. The flatness  of nanographene would be disrupted, leading to corrugation in the graphitic $sp^2$ structure. Therefore, it is essential to verify whether each PTM polymer suffers from steric hindrance. In DFT simulations, atomic forces can be calculated, and by monitoring the atomic displacements caused by these forces, the stable structure can be identified. This allows us to investigate the stable structure of the designed nanographene and assess whether it is affected by steric hindrance.

To calculate the optimal structure and self-consistent field (SCF), we used the norm-conserving pseudopotentials C.pbe-tm-gipaw.UPF and H.pbe-tm-gipaw.UPF in Quantum ESPRESSO.\cite{QE-1, QE-2, QE-3} The calculations were performed at the $\Gamma$ point for a unit cell with a vacuum layer having a width of approximately  15 \AA. The kinetic energy cutoff was set to 30 Ry for wave functions and 300 Ry for the charge density and potential.

The norm-conserving pseudopotentials that we used produced band structures of graphene that were indistinguishable from those obtained using other norm-conserving potentials. Therefore, these pseudopotentials can be used in the calculations.

The material structure was considered stable when the optimization procedure converged to a result. The lateral size of the simulation cell was fixed into a rhombohedral shape in the planar directions. Consequently, each obtained structure was subjected to small internal stresses. Given that our primary goal was to establish a systematic method for designing these materials, we used slightly stressed structures. The DFT simulations corroborated the presence of VANG in a supercell and the stability of the structures shown in Fig.~\ref{f2}.   Additionally,  the attained stable structure confirmed its planarity, indicating that the designed nanographene is not compromised by steric hindrance. When on-top hydrogen adsorption was used, the planar structure underwent slight deformation owing to the formation of $sp^3$ type bonds between the carbon and adsorbed hydrogen. The surface hydrogen distances between PTMs were obtained as follows: in VANG-A, the distance is $l_A = 5.97 \AA$, while in VANG-B, the vertical hydrogen distance is $l_{B,v} = 9.75 \AA$ and the diagonal hydrogen distance is $l_{B,d} = 12.42 \AA$.

\subsection{Electronic Structures of Polymerized Vacancy-created Armchair Nanographene}

\begin{figure}[h]
    \centering
    \includegraphics[width=8cm]{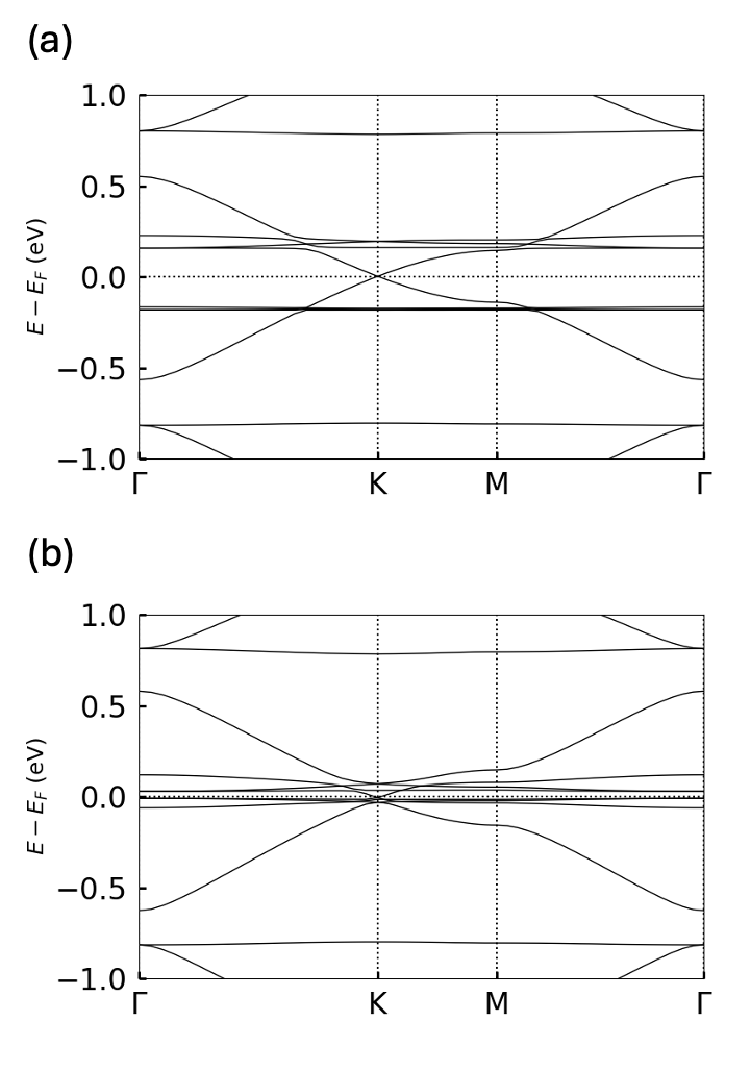}
    \caption{Band structure of (a) VANG-A nanographene and (b) VANG-B nanographene calculated by DFT}
    \label{f5}
\end{figure}

Following the modeling method described in the previous section, we obtained molecular models of VANG-A and VANG-B, for which we calculated the electronic band structure using DFT.  Fig.~\ref{f5} presents the calculated bands of these structures, where the Fermi energy $E_F$ is set as the energy origin. Notably,  the DFT band dispersion exhibits qualitative agreement with the so-TBM results. As intended, six nearly flat bands are observed. Although the energy splitting of the flat bands near the Fermi energy is evident, especially for VANG-A, the DFT band structure can be largely reproduced by incorporating weak second- and third-nearest neighbor transfers into the so-TBM. The bands exhibiting Dirac dispersion are nearly identical to the so-TBM bands in appearance, with only minimal hybridization with the flat bands.

In VANG-A, bands corresponding to localized zero modes split into two groups of three within $\pm0.21$ eV of $E_F$, exhibiting an energy difference of 0.4081 eV. In the so-TBM, the two localized zero modes that appear in adjacent PTMs are disjoint from each other and have no overlapping sites. However, as shown in the next section, the primary cause of the flat band splitting observed in the DFT band is the weak hybridization that occurs between these disjoint zero modes. 

In VANG-B, bands corresponding to localized zero modes split into two groups of three within $\pm 0.1$ eV of $E_F$, exhibiting an energy difference of 0.1803 eV. Once again, the primary cause of the split is the weak hybridization that occurs between the disjoint zero modes. The distinction between VANG-A and VANG-B is not confined to the energy scale of the split. As we will discuss later, the long-range transfer between these modes is also slightly different. 

We adopted the Wannierization method using the Wannier90 package for our nanographene structures.\cite{wannier90-1, wannier90-2, wanier90-3} A group of eight spectral bands around $E_F$ is flanked by gaps at the top and bottom, as shown in Figs.~\ref{f5}(a) and \ref{f5}(b). This feature was also observed in the bands of the so-TBM, as shown in Fig.~\ref{f3}. To capture these features, the upper and lower energy windows above and below the Fermi energy ($E_F$ ) were adjusted to include eight bands for both VANG-A and VANG-B. Therefore, the Wannierization used here encompasses eight bands in the vicinity of the Fermi energy, including six bands with localized zero modes and two bands with Dirac dispersions.

The initial values of the Wannier functions were set using $p_z$ orbitals on carbon atoms that bonded with the surface on-top hydrogen and carbon atoms either one level above or one level below the central carbon atom in each PTM. This setting effectively captures the two Dirac-like modes included in the low-lying branches considered for each system. In addition to the two Wannier orbitals that yield the Dirac dispersion, we obtained six orbitals corresponding to the localized zero modes

\begin{figure}[h]
    \centering
    \includegraphics[width=8cm]{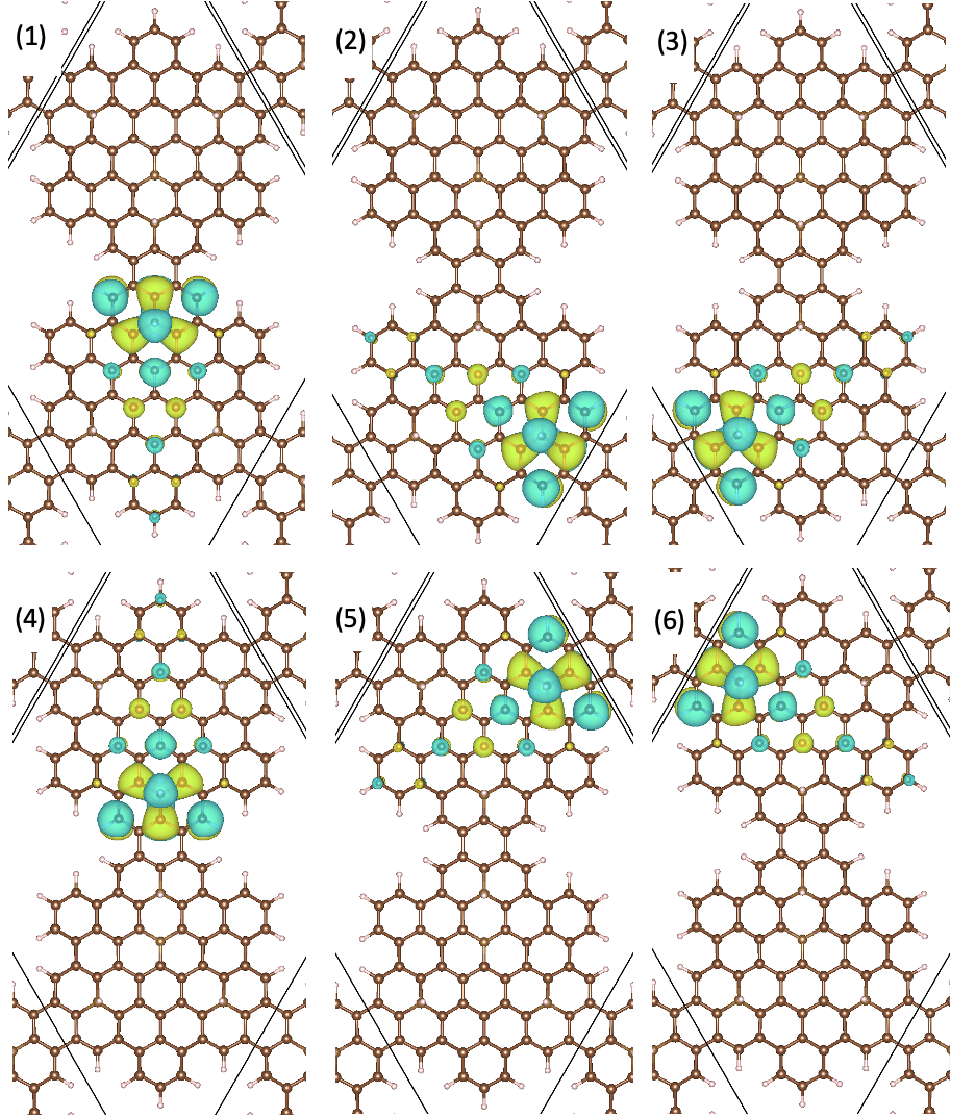}
    \caption{(Color online) Real-space Wannier functions of the localized zero modes in VANG-A nanographene }
    \label{f6}
\end{figure}

\begin{figure}[h]
    \centering
    \includegraphics[width=8cm]{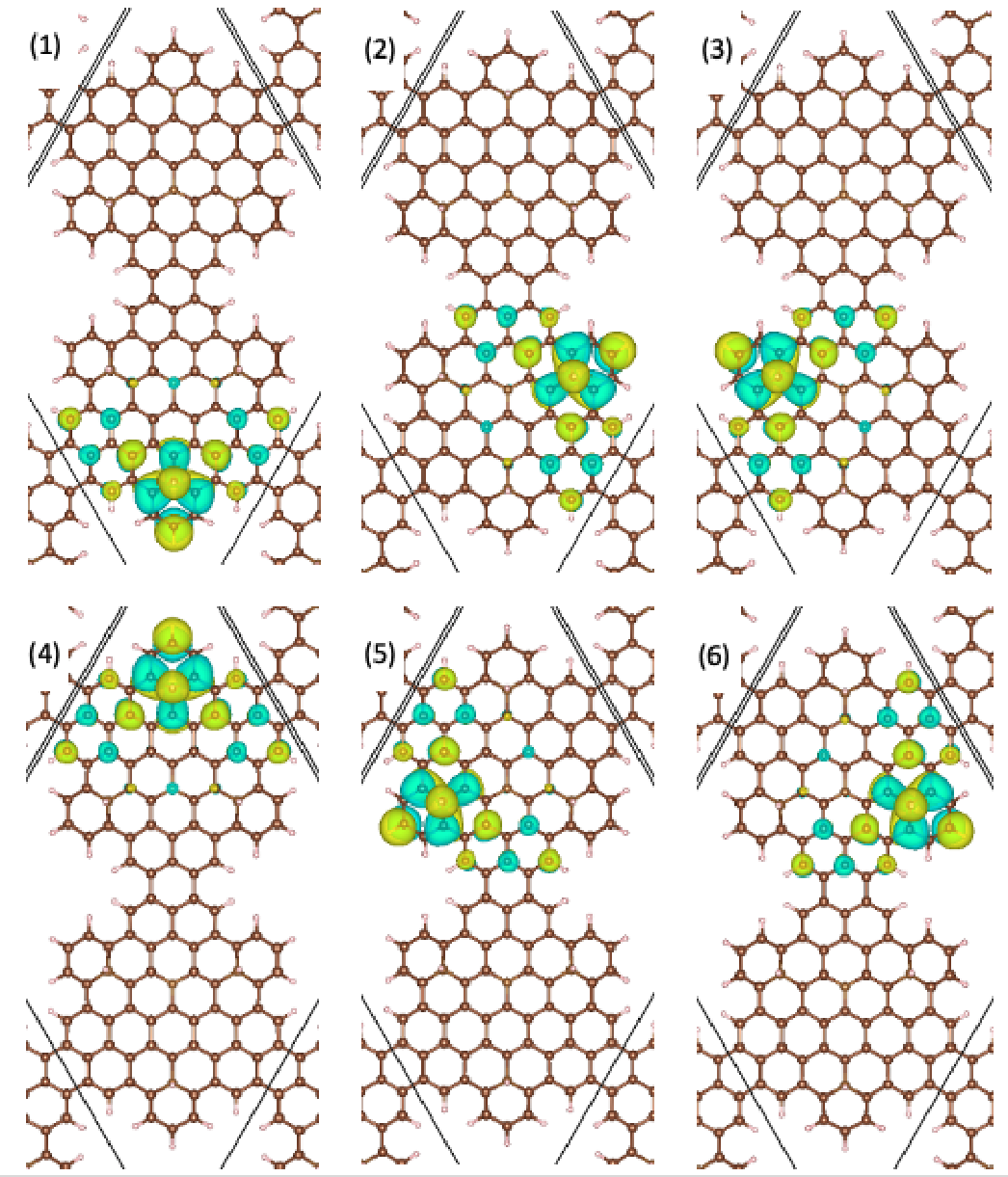}
    \caption{(Color online) Real-space Wannier functions of the localized zero modes in VANG-B nanographene }
    \label{f7}
\end{figure}

Figs.~\ref{f6} and \ref{f7} show the real-space Wannier functions for VANG-A and VANG-B, respectively. In these six orbitals, the wave function amplitude is non-zero only at A-sites and alternates with sign reversal. Each of the six wavefunctions has a node at each B-site. Therefore, the orbitals in the flat bands are characterized as NBMOs. These flat bands and their corresponding Wannier functions are primarily generated by the zero modes in the so-TBM. Therefore, these orbitals correspond to the zero modes of the so-TBM. In fact, the distribution of the wave function amplitudes in Fig.~\ref{f6} closely resembles the spatial distribution of the corresponding eigenzero mode of the so-TBM. The orbital energy is slightly shifted by 2 meV from the zero energy, {\it i.e}., the Fermi level, owing to transfer interactions between third-neighboring sites and beyond.

The degeneracy in the orbital energy of these six modes should be mostly preserved owing to the $C_3$ symmetry around the center of each PTM. Indeed, these six wavefunctions agree in the orbital energy in an error range of 0.5 meV. However, despite the structural similarity between VANG-A and VANG-B, there are considerable differences in band dispersion, particularly in the magnitude of the energy splitting. In the next subsection, we will demonstrate that this variation originates from the distance between the zero modes and their mutual arrangement.


\subsubsection{Analysis of the Hopping Hamiltonian}

Because the DFT simulations identify a flat band and the six orbitals of its Wannier functions are identified as NBMOs corresponding to the zero modes of the so-TBM, we can use these orbitals to construct an effective hopping Hamiltonian. The appearance of a non-zero bandwidth and even an energy splitting in the flat bands for both VANG-A and VANG-B yields a non-trivial low energy hopping Hamiltonian $H_{\rm Wan-TBM}$. 

We introduce some indices to specify the Wannier orbitals. Let the position vector of a unit cell be given by the 2D lattice vectors ${\bf a}_1$ and ${\bf a}_2$ (Fig.~\ref{f1}) as ${\bf R} = m_1{\bf a}_1 + m_2{\bf a}_2$. The indices $m_1$ and $m_2$ specify the unit cell. The two integers $m_1$ and $m_2$ are in the range $0\le m_j < M$ for $j=1,2$, where $M$ is a large integer. The six Wannier orbitals were labeled as $\varphi_{m_1,m_2,i}({\bf r})$ with $i=1,2,\cdots,6$. Typically, the orbital in Fig.~\ref{f6}(1) is $\varphi_{0,0,1}({\bf r})$.

The Wannierization yields the hopping Hamiltonian by evaluating matrix elements of $t_{ij}^{m_1m_2} = \langle \varphi_{m_1,m_2,j} | H_{\rm DFT-eff} | \varphi_{0,0,i} \rangle$. The DFT effective Hamiltonian, $H_{\rm DFT-eff}$, is determined via self-consistent simulation of the electron charge density. We introduced the fully indexed transfer parameters $t_{ij}^{m_1m_2}$ to represent the transfer from $\varphi_{0,0,i}({\bf r})$ to $\varphi_{m_1,m_2,j}({\bf r})$. Owing to the symmetry of the system, we found that many parameters were mutually consistent and could be expressed by a small number of parameters. 

Table~II summarizes the hopping terms obtained by Wannierization. We refer to the hopping terms among the three zero modes in one PTM as the intra-PTM transfer. For example, $\bar{t}=0.0135$ eV for VANG-A represents this intra-PTM transfer parameter and is identical to $t_{12}^{00}$, which represents electron hopping from $\varphi_{0,0,1}({\bf r})$ to $\varphi_{0,0,2}({\bf r})$. The value is an order of magnitude smaller than that of the largest inter-PTM hopping $t=-0.1788$ eV which is identical to $t_{14}^{00}$. The degeneracy in the orbital energy should be mostly preserved owing to the $C_3$ symmetry around the center of each PTM, which is reflected in the hopping parameters. 

Let $H_{\rm intra-PTM-ZM}$ represent the hopping terms among the three zero modes in one PTM and $H_{\rm DM}$ represent the hopping terms among the Dirac-like modes. We also introduce $H_{\rm ZM-DM}$ representing the hybridization terms between zero modes and Dirac-like modes in the 8 excitation branches. Then, we can separately define the inter-PTM hopping term $H_{\rm inter-PTM-ZM}$, which represents the hopping terms connecting a zero mode in a specified PTM, {\it i.e.}, $\varphi_{0,0,i}({\bf r})$, to zero modes in the other PTMs. The total hopping Hamiltonian $H_{\rm Wan-TBM}$ is then written as,  
\begin{eqnarray} H_{\rm Wan-TBM} &=& H_{\rm intra-PTM-ZM} + H_{\rm inter-PTM-ZM} \nonumber \\ && + H_{\rm ZM-DM} + H_{\rm DM}. \end{eqnarray}
The simulation results showed that the term of $H_{\rm ZM-DM}$ is negligible because the transfer between any $\varphi_{0,0,i}({\bf r})$ and a Dirac mode is less than 0.2 meV. Therefore, we can separately treat electrons in the zero modes from the Dirac-like modes. 

Many inter-PTM hopping terms exhibit a decaying nature with increasing distance. We confirmed that long-range hopping terms connecting a zero mode farther than the nearest-neighbor sites to another zero mode were negligible. By neglecting these long-range transfer terms other than $t$, $H_{\rm inter-PTM-ZM}$ is simplified. We call it $H_t$. We can safely focus on $H_{\rm intra-PTM-ZM}$ given by $\bar{t}$ and $H_t$ given by $t$. 

Here, we note the importance of long-range hopping between $\pi$-orbitals in deriving an effective Hamiltonian. First, these Wannier orbitals closely resemble the zero modes of the so-TBM in terms of spatial spread and probability amplitude distribution. Therefore, they maintain the disjoint orbital property of so-TBM, and the nodes of the wave function are on the carbon atoms at the joints of adjacent PTMs. Consequently, if we consider only a hopping term between adjacent $\pi$ orbitals, the contribution to t disappears.  

$t$ exceeding 10 meV is caused by longer-distance inter-$\pi$-orbital hopping. Other studies \cite{doi:10.1021/acs.nanolett.9b01773,PhysRevB.108.155423} have highlighted the importance of third-nearest $\pi$-orbital hopping in zero-mode hopping. However, it is also true that $t$ effectively incorporates hopping contributions between $\pi$-orbitals over distances longer than those between third-nearest $\pi$-orbitals. The Wannierization method appropriately captures this inter-zero-mode hopping term $t$. Our method is based on DFT and efficiently derives a low-energy effective Hamiltonian. 

When $H_{\rm inter-PTM-ZM}$ or $H_t$ induces local hybridization between a pair of neighboring zero modes, the flat bands split into bonding and antibonding bands but maintain their flatness. In VANG-A, the inter-NBMO hybridization by $t=t_{14}^{00}$ led to the formation of bonding and antibonding states constructed from $\varphi_{m_1,m_2,1}({\bf r})$ and $\varphi_{m_1,m_2,4}({\bf r})$ at each ($m_1$,$m_2$)-th unit cell. If $\bar{t}$ is set to zero, these bonding and antibonding modes become the eigenstates of $H_{Wan-TBM}$ . Even when $t=-0.1788$ eV for all inter-PTM hopping terms, the bonding (and antibonding) states remained disjointed, and no interlinking occurred between these bonding states. Consequently, we obtained two groups of flat bands at  $\pm 0.1788$ eV. These results correspond to the splitting of the central six flat bands into bonding and antibonding bands.

Adding a finite $\bar{t}$ slightly disperses these flat bands, explaining the results in Fig.~\ref{f5}(a). Preserving the band's flatness ensures that NBMO-to-NBMO transfer over long distances is negligible enough to justify the model's evaluation. The following section justifies our modeling by contrasting the energy scale of $\bar{t}$ with the inter-zero-mode direct exchange interactions.

\begin{table}
    \centering
    \begin{tabular}{lrrrrrrrr} \hline
   Hopping term & & & & & & &  Value[eV] \\ \hline
   VANG-A \\
    $\bar{t} $& & & & & & &  0.0135 \\
    $t $& & & & & & & -0.1788\\
   \hline
   VANG-B \\
    $\bar{t} $& & & & & & & 0.0055 \\
    $t_v $ & & & & & & & -0.0214\\
    $t_d $ & & & & & & & -0.0234\\
   \hline
   \end{tabular}
\label{t1}
\caption{Evaluation of tight-binding parameters. $t$ and $\bar{t}$ are the inter-PTM and intra-PTM transfer integrals, respectively. In VANG-B, $t_v$ and $t_d$ represent vertical hopping ({\it e.g.}, $t_{26}^{00}$), and diagonal hopping ({\it e.g.}, $t_{25}^{00}$), respectively. 
}
\end{table}

DFT simulations revealed differences between VANG-A and VANG-B band structures that were not evident in so-TBM. Notably, in Fig.~\ref{f5}, VANG-A's localized zero-mode bands exhibited a clear split while maintaining flatness, unlike VANG-B's less split and less flat localized zero-mode bands. Wannierization revealed greater inter-PTM transfer between localized zero modes for VANG-A, which also had a larger energy difference between localized zero mode bands compared to VANG-B. VANG-A's enhanced localized zero-mode transfer can be attributed to the closer surface hydrogen distance in each inter-PTM. This implies a greater overlap of localized zero modes in VANG-A compared to VANG-B. The energy difference between zero mode bands is influenced by the magnitude of inter-PTM transfer. Therefore, increasing inter-PTM transfer can widen the energy gap between localized zero mode bands.

In VANG-B, even if intra-PTM transfer is neglected as in VANG-A, inter-PTM transfer still interconnects the entire periodic structure of zero modes. This hinders electron localization, leading to the dispersion and bending of the flat bands (Fig.~\ref{f5}(b)).

\subsubsection{Evaluation of Inter-electron Interaction Terms}

We used the Pariser--Parr--Pople (PPP) method\cite{PPP} to calculate the electronic Coulomb interaction $U$ and direct exchange interaction $J$ for VANG-A and VANG-B. These parameters were specifically evaluated for the zero modes, not the $\pi$ orbitals. The PPP method uses an approximate approach for calculating interaction integrals. To ensure that the long-range Coulomb repulsion behaved inversely proportional to the distance $R$ as $1/R$, we approximated the integral kernel as $1/(1/U_0+R)$, where $U_0$ denotes the onsite Coulomb interaction at a $\pi$ orbital of graphene. The value of $U_0$ was set to 9.3 eV, obtained from DFT and constrained random-phase approximation (cRPA) calculations of graphene. \cite{PhysRevLett.106.236805} For simplicity, we assumed that the wave function amplitudes of the Wannierization results at each carbon site could be approximated by the corresponding zero mode of the so-TBM. Table III presents the calculation results, where $U$, $U'$, and $U''$ represent the onsite, intra-PTM, and inter-PTM Coulomb interactions, respectively ({\it e.g.}, $U_1$, $U^{00}_{12}$, and $U^{00}_{14}$). Here, $U_i$ denotes the intra-orbital Coulomb interaction in the $i$th Wannier orbital while $U^{m_1m_2}_{ij}$ indicates the inter-orbital Coulomb interaction between $\varphi_{0,0,i}({\bf r})$ and $\varphi_{m_1,m_2,j}({\bf r})$ . $J^{m_1m_2}_{ij}$ represents the direct exchange interaction between $\varphi_{0,0,i}({\bf r})$ and $\varphi_{m_1,m_2,j}({\bf r})$. The value of $J$ represents direct exchange interactions in PTMs, corresponding to $J^{00}_{12}$.

\begin{table}
    \centering
    \begin{tabular}{lrrrrrr} \hline
   Interaction term & & & & & &   Value[eV] \\ \hline
   VANG-A \\
   $U $& & & & & &  3.829 \\
   $U' $& & & & & & 1.681 \\
   $U'' $& & & & & & 1.994 \\
   $J $& & & &  & & 0.031 \\
   \hline
   VANG-B \\
   $U $& & & & & & 3.597 \\
   $U'$& & & & & & 1.678 \\
   $U_v''$& & & & & & 1.267 \\
   $U_d''$& & & & & & 1.075 \\
   $J $& & & & & & 0.043 \\
   \hline
   \end{tabular}
\label{t2}
\caption{ Coulomb interaction $U$ and direct exchange interaction $J$ for VANG-A and VANG-B. $U$ represents the onsite Coulomb interaction at a localized zero mode. $U'$ represents the Coulomb interaction between zero modes within one PTM (intra-PTM interaction), and $U''$ represents the Coulomb interaction between the closest pair of zero modes in neighboring PTMs (inter-PTM interaction). Additionally, in VANG-B, $ U_v''  $ and $ U_d'' $ represent vertical Coulomb interaction ({\it e.g.}, $U_{26}^{00}$), and diagonal Coulomb interaction ({\it e.g.}, $U_{25}^{00}$), respectively. $J$ represents the direct ferromagnetic exchange between Wannier orbitals within one PTM (intra-PTM).}
\end{table}

\subsubsection{Effective Spin Hamiltonian}

Based on the abovementioned results, we derived an extended Hubbard Hamiltonian for VANG-A. We used the Wannier orbitals $\varphi_{m_1,m_2,i}({\bf r})$ to define the creation and annihilation operators $c_{m_1,m_2,i,\sigma}^\dagger$ and $c_{m_1,m_2,i,\sigma}$, respectively, for the $i$-th zero modes ($i=1, 2, \cdots, 6$), where $\sigma=\uparrow, \downarrow$ denotes the electron spin. Neglecting hopping terms with transfer parameters smaller than $t$, the transfer Hamiltonian can be expressed as a sum of three primary hopping terms:

\begin{eqnarray}
H_t &=& t \sum_{m_1,m_2} \sum_\sigma \left( 
c_{m_1,m_2,1,\sigma}^\dagger c_{m_1,m_2,4,\sigma} 
+ {\rm H.c.}
\right) \nonumber \\
&+& t \sum_{m_1,m_2} \sum_\sigma \left( 
c_{m_1+1,m_2,3,\sigma}^\dagger c_{m_1,m_2,5,\sigma} 
+ {\rm H.c.}
\right) \nonumber \\
&+& t \sum_{m_1,m_2} \sum_\sigma \left( 
c_{m_1,m_2+1,2,\sigma}^\dagger c_{m_1,m_2,6,\sigma} 
+ {\rm H.c.}
\right). 
\end{eqnarray}
Excluding a constant term, the following form is obtained as an extended diagonal Hubbard interaction term:

\begin{eqnarray}
H_{U} 
&=& \frac{U}{2} \sum_{m_1,m_2} \sum_{i=1}^6 : 
\left(n_{m_1,m_2,i,\uparrow} + n_{m_1,m_2,i,\downarrow} - 1 \right)^2 : 
\nonumber \\
&+& U' \sum_{m_1,m_2} \sum_{\sigma,\sigma'} 
\sum_{\langle i, j \rangle \in S_1} \left( 
n_{m_1,m_2,i,\sigma} n_{m_1,m_2,j,\sigma'} 
\right) \nonumber \\
&+& U'' \sum_{m_1,m_2} \sum_{\sigma,\sigma'} 
\left( 
n_{m_1,m_2,1,\sigma} n_{m_1,m_2,4,\sigma'} 
\right. \nonumber \\
&& 
+ \left. 
n_{m_1+1,m_2,3,\sigma} n_{m_1,m_2,5,\sigma'} 
+ n_{m_1,m_2+1,2,\sigma} n_{m_1,m_2,6,\sigma'} 
\right)  \nonumber \\
&+&  {\rm const.} 
\end{eqnarray}
The number operator $n_{m_1,m_2,i,\sigma}$ is defined as $c^\dagger_{m_1,m_2,i,\sigma}c_{m_1,m_2,i,\sigma}$. Here, $:A:$ represents the normal-ordered product of the operator $A$ with respect to the creation and annihilation operators, where the order of operators is exchanged using the anticommutation property, ensuring that annihilation operators are placed to the left of the creation operators.  
The set of site pairs, $S_1$, 
contains $\langle i, j \rangle =$ $\langle 1,2 \rangle$, 
$\langle 2,3 \rangle $, $\langle 3,1 \rangle $, 
$\langle 4,5 \rangle $, $\langle 5,6 \rangle $, 
$\langle 6,4 \rangle $. 
n a PTM, an inter-zero-mode exchange interaction arises, denoted here as $H_J$. 
\begin{eqnarray}
H_J
&=&
-2J \sum_{m_1,m_2} \sum_{\langle i, j \rangle \in S_1}
\left\{ \boldsymbol{S}_{m_1,m_2,i} \cdot \boldsymbol{S}_{m_1,m_2,j} \right.
\nonumber \\
&+ & \left. \frac{1}{4}(n_{m_1,m_2,i,\uparrow}+n_{m_1,m_2,i,\downarrow})(n_{m_1,m_2,j,\uparrow}+n_{m_1,m_2,j,\downarrow})\right\} . \nonumber \\
\end{eqnarray}
where $\boldsymbol{S}_{m_1,m_2,i}$ indicates an S=1/2 spin operator on the Wannier orbital $\varphi_{m_1,m_2,i}({\bf r})$.

The total Hamiltonian for VANG-A is given as 
\begin{equation}
    H_A = H_U + H_J + H_t.
\end{equation}

In this case,  $U/t$ becomes 21 for VANG-A. ndicating that the system resides in the strong-correlation regime of the extended Hubbard model. Consequently, we can treat $H_U+H_J \equiv H_0$ as the non-perturbation terms and $H_t \equiv V$ as the perturbation term. The half-filled state of $H_0$ corresponds to an array of $S=3/2$ spins on a honeycomb lattice.  We used the second-order Rayleigh--Schr{\"o}dinger perturbation theory to the Hubbard Hamiltonian $H$.\cite{Lindgren} The resulting low-energy effective Hamiltonian is given by:

\begin{equation}
    H_{A \; eff}^{(2)} = \sum_{<m,n>} \mathcal{J}_A \left(\boldsymbol{S}^{(\frac{3}{2})}_{\alpha_m} \cdot \boldsymbol{S}^{(\frac{3}{2})}_{\beta_n} - \frac{9}{4}\right), \mathcal{J}_A = \frac{4t^2}{9(U - U'' +2J)}.
    \label{Qterm_A}
\end{equation}

Here, $\boldsymbol{S}^{(\frac{3}{2})}_{\alpha_m(\beta_n)}$ represents the spin operator with $S = 3/2$ on the $m$-th $\alpha$-PTM ($n$-th $\beta$-PTM). Using the values obtained in Tables II and III, the value of $\mathcal{J}_A$ was calculated as 8 meV. Subsequently, the effective Hamiltonian transformed into an antiferromagnetic $S=3/2$ honeycomb Heisenberg model.

Analogous to VANG-A, we constructed the Hubbard Hamiltonian for VANG-B.  By applying the second-order Rayleigh--Schr{\"o}dinger perturbative expansion, the low-energy effective Hamiltonian is derived as

\begin{equation}
\begin{split}
   & H_{B \; eff}^{(2)} = \sum_{<m,n>} \mathcal{J}_B \left(\boldsymbol{S}^{(\frac{3}{2})}_{\alpha_m} \cdot \boldsymbol{S}^{(\frac{3}{2})}_{\beta_n} - \frac{9}{4}\right),\\
    &\mathcal{J}_B = 2 \times \frac{4t_v^2}{9(U - U_v'' +2J) } + 2 \times \frac{4t_d^2}{9(U - U_d'' +2J) }.
    \label{Qterm_B}
\end{split}
\end{equation}

Using the results obtained in Sections 3.2.1 and 3.2.2, the value of $\mathcal{J}_B$ can be calculated as 0.4 meV. Therefore, we conclude that the antiferromagnetic S=3/2 honeycomb Heisenberg model is also realized in VANG-B nanographene. However, stronger effective antiferromagnetic interactions are observed in VANG-A nanographene. This disparity is attributed to the larger inter-PTM hopping values in VANG-A. In essence, to achieve stronger effective antiferromagnetic interactions in the design of nanographene using PTM, it is crucial to design nanographene such that the distance between the surface hydrogens of adjacent PTMs is minimized, thereby promoting stronger inter-PTM hopping.

The proposed 2D structures based on molecular design exhibited considerable deviations from the behavior predicted by the so-TBM. This discrepancy can be attributed to the presence of a band structure with a split between localized zero modes. Guided by the design rules, various carbon skeletal structures with localized zero modes can be proposed. For instance, a molecular model with a larger carbon array would allow for more freedom in selecting surface hydrogen positions compared to nanographene structures such as VANG-A. This flexibility would enable more selective manipulation of electronic interactions and transfers. By manipulating localized magnetic arrangements and interactions, molecules capable of realizing not only the honeycomb Heisenberg model but also a broader range of spin systems can be created.

Incorporating terms up to the fourth order in the RS perturbation introduces various terms, including the bi-quadratic term (BQt).\cite{Lindgren, PhysRevB.101.024418} However, BQt is relatively small as a higher-order term compared to the quadratic term in Eq.~(\ref{Qterm_A}). Strategies and methods for enhancing BQt are discussed in detail elsewhere.

\section{Conclusions}
We developed a design methodology that generates a molecular framework in which localized electron wave functions emerge spontaneously. Beyond the emergence of zero modes, magnetic inter-electron interactions can be engineered according to the design. The design method was validated through DFT simulations, confirming the configuration of localized zero modes. The low-energy electron Hamiltonian obtained via the Wannierization method naturally defines a transfer Hamiltonian. We elucidate the mechanism by which inter-electron magnetic interactions lead to the emergence of the spin Hamiltonian as the low-energy effective Hamiltonian.

We present several illustrative examples of 2D nanographene structures with localized zero modes constructed using our proposed design method. DFT simulations and Wannierization results for these nanographene structures reveal changes in the band structure around localized zero modes, which can be attributed to distinctive variations in transfer characteristics. Our approach uses PTMs with multiple spatially overlapping zero modes, enabling molecular designs that generate enhanced magnetic interactions beyond the simple placement of zero modes as isolated NBMOs. The ferromagnetic interaction terms that give rise to large spin moments of $S>1$ are attributed to spin alignment induced by direct exchange interactions because multiple zero modes in a PTM yield degenerate orthogonal molecular orbitals with spatial overlap. The effective interaction mediating these localized spin moments with antiferromagnetic interactions arises from the weak electron transfer between zero modes in neighboring PTMs. Our method leads to a Hubbard U term that is several orders of magnitude larger than the inter-PTM transfer for zero modes. Consequently, the electron system defined by the zero-mode array enters a genuine strongly correlated regime, signifying the formation of a stable localized spin state. By considering electronic correlation effects and electron transfers, we were able to design magnetic structures. The tunability of design and transfer parameters in our method highlights the potential of molecular design as a powerful tool for engineering materials with diverse magnetic structures, extending beyond those presented in this study.
\\

\section*{Acknowledgement}
We would like to thank Yasuhiro Oishi for the helpful discussions. We acknowledge the financial support by Grant-in-Aids for Scientific Research (JSPS KAKENHI) Grants No. JP22K04864, No. JP21K13887, No. JP23H03817, and No. JP24K17014. Computations in this work were made using the facilities of the Supercomputer Center, the Institute for Solid State Physics, the University of Tokyo, the University of Kyushu.

\bibliographystyle{jpsj}
\bibliography{main}

\begin{thebibliography}{10}

\bibitem{PhysRevLett.86.5188}
R.~Raussendorf and H.~J. Briegel: Phys. Rev. Lett. {\bfseries 86} (2001) 5188.

\bibitem{PhysRevLett.93.040503}
M.~A. Nielsen: Phys. Rev. Lett. {\bfseries 93} (2004) 040503.

\bibitem{PhysRevLett.97.110501}
N.~C. Menicucci, P.~van Loock, M.~Gu, C.~Weedbrook, T.~C. Ralph, and M.~A. Nielsen: Phys. Rev. Lett. {\bfseries 97} (2006) 110501.

\bibitem{PhysRevLett.103.240504}
D.~N. Biggerstaff, R.~Kaltenbaek, D.~R. Hamel, G.~Weihs, T.~Rudolph, and K.~J. Resch: Phys. Rev. Lett. {\bfseries 103} (2009) 240504.

\bibitem{PhysRevLett.105.013902}
S.~L\'opez-Aguayo, Y.~V. Kartashov, V.~A. Vysloukh, and L.~Torner: Phys. Rev. Lett. {\bfseries 105} (2010) 013902.

\bibitem{PhysRevLett.105.093601}
S.~E. Economou, N.~Lindner, and T.~Rudolph: Phys. Rev. Lett. {\bfseries 105} (2010) 093601.

\bibitem{PhysRevLett.105.110502}
S.~D. Bartlett, G.~K. Brennen, A.~Miyake, and J.~M. Renes: Phys. Rev. Lett. {\bfseries 105} (2010) 110502.

\bibitem{PhysRevLett.106.070501}
T.-C. Wei, I.~Affleck, and R.~Raussendorf: Phys. Rev. Lett. {\bfseries 106} (2011) 070501.

\bibitem{PhysRevLett.114.247204}
M.~Koch-Janusz, D.~I. Khomskii, and E.~Sela: Phys. Rev. Lett. {\bfseries 114} (2015) 247204.

\bibitem{Morishita2021}
N.~Morishita, Y.~Oishi, T.~Yamaguchi, and K.~Kusakabe: Applied Physics Express {\bfseries 14} (2021) 121005.

\bibitem{PhysRevB.108.155423}
J.~C.~G. Henriques and J.~Fern\'andez-Rossier: Phys. Rev. B {\bfseries 108} (2023) 155423.

\bibitem{doi:10.1021/acs.nanolett.3c04915}
J.~Henriques, M.~Ferri-Cort^^c3^^a9s, and J.~Fern^^c3^^a1ndez-Rossier: Nano Letters {\bfseries 24} (2024) 3355.

\bibitem{doi:10.1021/ja00456a010}
W.~T. Borden and E.~R. Davidson: Journal of the American Chemical Society {\bfseries 99} (1977) 4587.

\bibitem{Shima-Aoki}
N.~Shima and H.~Aoki: Phys. Rev. Lett. {\bfseries 71} (1993) 4389.

\bibitem{doi:10.1080/10587259308035713}
W.~T. Borden: Molecular Crystals and Liquid Crystals Science and Technology. Section A. Molecular Crystals and Liquid Crystals {\bfseries 232} (1993) 195.

\bibitem{Aoki-Imamura}
A.~I. Yuriko~Aoki: International Journal of Quantum Chemistry {\bfseries 74} (1999) 491.

\bibitem{MORISHITA_PLA_2021}
N.~Morishita and K.~Kusakabe: Physics Letters A {\bfseries 408} (2021) 127462.

\bibitem{doi:10.1021/acs.nanolett.9b01773}
R.~Ortiz, R.~A.~Boto, N.~Garc^^c3^^ada-Mart^^c3^^adnez, J.~C.~Sancho-Garc^^c3^^ada, M.~Melle-Franco, and J.~Fern^^c3^^a1ndez-Rossier: Nano Letters {\bfseries 19} (2019) 5991^^e2^^80^^935997.

\bibitem{Hohenberg-Kohn}
P.~Hohenberg and W.~Kohn: Phys. Rev. {\bfseries 136} (1964) B864.

\bibitem{Kohn-Sham}
W.~Kohn and L.~J. Sham: Phys. Rev. {\bfseries 140} (1965) A1133.

\bibitem{Marzari-Vanderbilt}
N.~Marzari and D.~Vanderbilt: Phys. Rev. B {\bfseries 56} (1997) 12847.

\bibitem{Enoki_Ando}
T.~Enoki and T.~Ando: {\em Physics and Chemistry of Graphene (Second Edition): Graphene to Nanographene} (Jenny Stanford Publishing, Singapole, 2019).

\bibitem{Krygowski}
M.~K. C.~E. T.~M.~Krygowski: {\em Aromaticity in Heterocyclic Compounds} (Springer-Verlag, Berlin, 2009).

\bibitem{Saito_Dresselhaus}
G.~D. R.~Saito and M.~S. Dresselhaus: {\em Physical properties of carbon nanotubes} (Imperial College Press, London, 1998).

\bibitem{Morishita2016}
N.~Morishita, G.~K. Sunnardianto, S.~Miyao, and K.~Kusakabe: Journal of the Physical Society of Japan {\bfseries 85} (2016) 084703.

\bibitem{Morishita2019}
N.~Morishita and K.~Kusakabe: Journal of the Physical Society of Japan {\bfseries 88} (2019) 124707.

\bibitem{SM}
(Supplemental material) The tight-binding model analysis of an additional electron site is provided online.

\bibitem{Lieb-Loss-McCann}
E.~H. Lieb, M.~Loss, and R.~J. McCann: Journal of Mathematical Physics {\bfseries 34} (1993) 891.

\bibitem{Lieb}
E.~H. Lieb: Phys. Rev. Lett. {\bfseries 62} (1989) 1201.

\bibitem{doi:10.1143/JPSJ.61.1165}
K.~Kusakabe and H.~Aoki: Journal of the Physical Society of Japan {\bfseries 61} (1992) 1165.

\bibitem{PhysRevLett.72.144}
K.~Kusakabe and H.~Aoki: Phys. Rev. Lett. {\bfseries 72} (1994) 144.

\bibitem{Ziatdinov}
M.~Ziatdinov, S.~Fujii, K.~Kusakabe, M.~Kiguchi, T.~Mori, and T.~Enoki: Phys. Rev. B {\bfseries 89} (2014) 155405.

\bibitem{QE-1}
P.~Giannozzi, O.~Baseggio, P.~Bonf^^c3^^a0, D.~Brunato, R.~Car, I.~Carnimeo, C.~Cavazzoni, S.~de~Gironcoli, P.~Delugas, F.~Ferrari~Ruffino, A.~Ferretti, N.~Marzari, I.~Timrov, A.~Urru, and S.~Baroni: The Journal of Chemical Physics {\bfseries 152} (2020) 154105.

\bibitem{QE-2}
P.~Giannozzi, O.~Andreussi, T.~Brumme, O.~Bunau, M.~B. Nardelli, M.~Calandra, R.~Car, C.~Cavazzoni, D.~Ceresoli, M.~Cococcioni, N.~Colonna, I.~Carnimeo, A.~D. Corso, S.~de~Gironcoli, P.~Delugas, R.~A. DiStasio, A.~Ferretti, A.~Floris, G.~Fratesi, G.~Fugallo, R.~Gebauer, U.~Gerstmann, F.~Giustino, T.~Gorni, J.~Jia, M.~Kawamura, H.-Y. Ko, A.~Kokalj, E.~K^^c3^^bc^^c3^^a7^^c3^^bckbenli, M.~Lazzeri, M.~Marsili, N.~Marzari, F.~Mauri, N.~L. Nguyen, H.-V. Nguyen, A.~O. de-la Roza, L.~Paulatto, S.~Ponc^^c3^^a9, D.~Rocca, R.~Sabatini, B.~Santra, M.~Schlipf, A.~P. Seitsonen, A.~Smogunov, I.~Timrov, T.~Thonhauser, P.~Umari, N.~Vast, X.~Wu, and S.~Baroni: Journal of Physics: Condensed Matter {\bfseries 29} (2017) 465901.

\bibitem{QE-3}
P.~Giannozzi, S.~Baroni, N.~Bonini, M.~Calandra, R.~Car, C.~Cavazzoni, D.~Ceresoli, G.~L. Chiarotti, M.~Cococcioni, I.~Dabo, A.~D. Corso, S.~de~Gironcoli, S.~Fabris, G.~Fratesi, R.~Gebauer, U.~Gerstmann, C.~Gougoussis, A.~Kokalj, M.~Lazzeri, L.~Martin-Samos, N.~Marzari, F.~Mauri, R.~Mazzarello, S.~Paolini, A.~Pasquarello, L.~Paulatto, C.~Sbraccia, S.~Scandolo, G.~Sclauzero, A.~P. Seitsonen, A.~Smogunov, P.~Umari, and R.~M. Wentzcovitch: Journal of Physics: Condensed Matter {\bfseries 21} (2009) 395502.

\bibitem{wannier90-1}
A.~A. Mostofi, J.~R. Yates, Y.-S. Lee, I.~Souza, D.~Vanderbilt, and N.~Marzari: Computer Physics Communications {\bfseries 178} (2008) 685.

\bibitem{wannier90-2}
A.~A. Mostofi, J.~R. Yates, G.~Pizzi, Y.-S. Lee, I.~Souza, D.~Vanderbilt, and N.~Marzari: Computer Physics Communications {\bfseries 185} (2014) 2309.

\bibitem{wanier90-3}
G.~Pizzi, V.~Vitale, R.~Arita, S.~Bl^^c3^^bcgel, F.~Freimuth, G.~G^^c3^^a9ranton, M.~Gibertini, D.~Gresch, C.~Johnson, T.~Koretsune, J.~Iba^^c3^^b1ez-Azpiroz, H.~Lee, J.-M. Lihm, D.~Marchand, A.~Marrazzo, Y.~Mokrousov, J.~I. Mustafa, Y.~Nohara, Y.~Nomura, L.~Paulatto, S.~Ponc^^c3^^a9, T.~Ponweiser, J.~Qiao, F.~Th^^c3^^b6le, S.~S. Tsirkin, M.~Wierzbowska, N.~Marzari, D.~Vanderbilt, I.~Souza, A.~A. Mostofi, and J.~R. Yates: Journal of Physics: Condensed Matter {\bfseries 32} (2020) 165902.

\bibitem{PPP}
K.~Nakada, M.~Igami, and M.~Fujita: Journal of the Physical Society of Japan {\bfseries 67} (1998) 2388.

\bibitem{PhysRevLett.106.236805}
T.~O. Wehling, E.~\ifmmode \mbox{\c{S}}\else \c{S}\fi{}a\ifmmode \mbox{\c{s}}\else \c{s}\fi{}\ifmmode \imath \else \i \fi{}o\ifmmode~\breve{g}\else \u{g}\fi{}lu, C.~Friedrich, A.~I. Lichtenstein, M.~I. Katsnelson, and S.~Bl\"ugel: Phys. Rev. Lett. {\bfseries 106} (2011) 236805.

\bibitem{Lindgren}
I.~Lindgren and J.~Morrison: {\em Atomic Many-Body Theory} (Springer-Verlag, Berlin, 1982).

\bibitem{PhysRevB.101.024418}
M.~Hoffmann and S.~Bl\"ugel: Phys. Rev. B {\bfseries 101} (2020) 024418.

\end{thebibliography}

\end{document}